\documentclass[a4paper, 12pt, sort&compress]{elsarticle}

\usepackage{amsmath}
\usepackage{amssymb}
\usepackage{subfig}
\usepackage{multirow}
\usepackage[english]{babel}
\usepackage{algorithm}
\usepackage{algpseudocode}
\usepackage{setspace}
\usepackage{xcolor}
\usepackage{hyperref}
\usepackage{cleveref}
\usepackage{bm}

\newcommand{\pderiv}[2]{\frac{\partial #1}{\partial #2}}
\newcommand{\bs}[1]{\boldsymbol{#1}}

\newcommand{\gammasD}{\gamma^s_{\text{D}}}
\newcommand{\gammasN}{\gamma^s_{\text{N}}}

\newcommand{\Igammafs}{\int_{\gamma^{f-s}}}

\newcommand{\T}{\; \mathrm{T}}

\newcommand{\myequal}{\, {=} \,}

\RequirePackage[margin=24mm]{geometry}

\begin{document}

\fontfamily{ptm}\selectfont

\begin{frontmatter}



\title{Insights into the performance of loosely-coupled FSI schemes based on Robin boundary conditions}

\author[]{Chennakesava Kadapa \corref{cor1}}
\ead{c.kadapa@bolton.ac.uk}

\address{
School of Engineering, University of Bolton, Bolton BL3 5AB, United Kingdom.}

\begin{abstract}
Robin boundary conditions are a natural consequence of employing Nitsche's method for imposing the kinematic velocity constraint at the fluid-solid interface. Loosely-coupled FSI schemes based on Dirichlet-Robin or Robin-Robin coupling have been demonstrated to improve the stability of such schemes with respect to added-mass. This paper aims to offer some numerical insights into the performance characteristics of such loosely-coupled FSI schemes based on Robin boundary conditions. Using numerical examples, we demonstrate that the improved stability due to the added damping term is actually at the expense of important dynamic characteristics of the structural sub-problem.
\end{abstract}

\end{frontmatter}

\graphicspath{{./figures/}}

\section{Introduction}
Numerical instabilities are inherent to any loosely-coupled numerical methods used for solving coupled problems. Added-mass instability is one such numerical instability encountered in the multiphysics problem of incompressible fluid-structure interaction (FSI). As per this instability, partitioned approaches in which the fluid-solid coupling is resolved by iterating over individual solvers for fluid and solid sub-problems have a stability limit with respect to the ratio of densities of solid and fluid domains, the thickness of the structure and flexibility of the structure. During the past two decades, significant efforts have been made into understanding these instabilities \cite{CausinCMAME2005,ForsterCMAME2007,BrummelenJAM2009,DettmerIJNME2013} as well as towards improving the stability limits \cite{BadiaJCP2008,NobileSIAMJSC2008,BurmanCMAME2009,GerardoSIAMNA2010,JaimanCMAME2011,FernandezCMAME2013,
BurmanIJNME2014FSI,DettmerIJNME2013,KadapaOE2020,BrummelenIJNMF2011,BanksJCP2014part1,
BanksJCP2014part2,SerinoJCP2019,GiganteCMA2021}.

Among the several techniques proposed in the literature towards improving the stability of loosely-coupled FSI schemes, the present work focuses on the novel approach proposed by Burman and Fernandez \cite{BurmanCMAME2009}. The approach presented in \cite{BurmanCMAME2009} combines Nitsche's method for enforcing the interface kinematic constraint and an interface pressure stabilisation to develop a loosely-coupled FSI scheme that is unconditionally stable. The Robin boundary condition, which is a natural consequence of employing Nitsche's method for weakly enforcing the velocity constraint at the interface, is shown to significantly enhance the stability of loosely-coupled FSI schemes \cite{BurmanCMAME2009,BurmanIJNME2014FSI}. Such a coupling has been successfully employed for simulating FSI problems involving thin, flexible structures \cite{RubergCMAME2012, RubergIJNMF2014}.

While partitioned schemes based on Dirichlet-Robin or Robin-Robin coupling certainly seem to have improved stability limits, to the best of the author's knowledge, the reasons for such behaviour have not been discussed so far in the literature. By ignoring the effects of interface pressure stabilisation, this paper aims to offer some insights into the performance of loosely-coupled FSI schemes based on Dirichlet-Robin coupling.

A close look at the treatment of Robin boundary conditions (BCs) presented in \cite{BurmanCMAME2009,BurmanIJNME2014FSI} indicates that the improved stability of explicit (or loosely-coupled) FSI schemes based on Robin BCs for the solid problem is not because of the boundary condition itself but rather due to how the solid velocity in the Robin BC is treated. To elaborate, the implicit treatment of solid velocity in the Robin BC as proposed in \cite{BurmanCMAME2009} introduces an artificial damping term into the structural model. This artificial damping term is believed to be responsible for the improved stability of the partitioned FSI scheme based on Robin BC for the solid problem. However, since the damping term is directly proportional to the penalty parameter in Nitsche's method, the amount of damping varies with the penalty parameter. Moreover, violation of the interface velocity constraint in the solid sub-problem even in the weak sense introduces a phase error in the structural response. The combined effect of added-damping and phase error becomes so significant for higher values of penalty parameter that it substantially alters the dynamic characteristics such as natural frequency and time period of the structural model. The response of the structure due to its altered dynamic characteristics is more pronounced for large time step sizes.

In this paper, we verify the above-stated shortcomings by considering two algorithms: one with the explicit treatment of solid velocity and the second with the implicit treatment.
For the demonstration, two numerical examples consisting of thick and thin beams in cross-flow in two dimensions are used. The numerical simulation framework used for this study is the cut-cell based methodology studied extensively in \cite{DettmerCMAME2016,KadapaCMAME2017rigid,KadapaCMAME2018,KadapaOE2020}. The governing equations and the formulation are presented in Section \ref{sec-goveqns}. The coupling strategy and its temporal discretisation are discussed in  Section \ref{sec-formulation}. The numerical examples are adapted for demonstrations in the present work are presented in Section  \ref{sec-examples}. The paper is concluded with Section \ref{sec-conclusions} with a summary of observations and conclusions drawn.

\section{Governing equations} \label{sec-goveqns}
This section summarises the governing equations for the fluid-structure interaction problems involving laminar, viscous, incompressible flows and flexible solids. The domain and boundary of the solid problem are denoted as $\Omega^s$ and $\Gamma^s$, respectively, in the original configuration, and as $\omega^s$ and $\gamma^s$, respectively, in the current configuration. The corresponding entities for the fluid domain are denoted, respectively, as $\Omega^f$ and $\Gamma^f$. The interface between the fluid and solid domains is denoted as $\Gamma^{f-s}$ in the original configuration and as $\gamma^{f-s}$ in the current configuration.

\subsection{Governing equations for the fluid problem}
For isothermal, laminar, viscous and incompressible flows, the Navier-Stokes equations can be written as,

\begin{subequations} \label{eqns-NS-main}
\begin{align}
\rho^f \pderiv{\bm{v}^f}{t} + \rho^f (\bm{v}^f \cdot \bm{\nabla}) \bm{v}^f - \nabla \cdot \bs{\sigma}^f &= \bm{f}^f && \mathrm{in} \quad \Omega^f,  \\
\bm{\nabla} \cdot \bm{v}^f &= 0 && \mathrm{in} \quad \Omega^f, \\
\bm{v}^f &= \bar{\bm{v}}^f && \mathrm{on} \quad \Gamma^f_D, \\
\bm{\sigma}^f \cdot \bm{n}^f &= \bar{\bm{t}}^f &&   \mathrm{in} \quad \Gamma^f_N, \\
\bm{v}^f(t \myequal 0) &= \bm{v}^f_0  && \mathrm{in} \quad \Omega^f, \\
p^f(t \myequal 0) &= p^f_0  && \mathrm{in} \quad \Omega^f,
\end{align}
\end{subequations}
where, $t$ is the time variable, $\bm{\nabla}$ is the gradient operator, $\bm{v}^f$ is the velocity of the fluid, $p^f$ is the pressure field in the fluid domain, $\rho^f$ is the density of the fluid, $\bm{f}^f$ is the body force on the fluid domain, $\bm{n}^f$ is the unit outward normal on the boundary $\Gamma^f$, $\bm{v}^f_0$ is the initial velocity, $p^f_0$ is the initial pressure, $\bar{\bm{v}}^f$ is the prescribed velocity on Dirichlet boundary $\Gamma^f_D$, and $\bar{\bm{t}}^f$ is applied traction on the Neumann boundary $\Gamma^f_N$. The stress $\bm{\sigma}^f$ employed in the present work is the pseudo-stress, and it is defined as,
\begin{align} \label{eqn-stress}
\bm{\sigma}^f := \mu^f \bm{\nabla} \bm{v}^f - p^f \, \bm{I},
\end{align}
where, $\bm{I}$ is the second-order identity tensor.

\subsection{Governing equations for the solid problem}
As the flexible solids considered in this work experience large deformations, finite strain formulation is employed to model the deformation behaviour of solid domains. The equations governing the elastodynamics can be written in the current configuration as,
\begin{align} \label{gov-eqns-solid}
\rho^s \pderiv{^2 \bm{d}^s}{t^2} - \nabla_{\bm{x}} \cdot \bs{\sigma}^s &= \bm{f}^s \quad\quad \mathrm{in} \quad \omega^s \\
\bm{d}^s &= \bar{\bm{d}}^s \quad\quad \mathrm{on} \quad \gammasD \\
\bm{\sigma}^s \cdot \bm{n}^s &= \bar{\bm{t}}^s \,\quad\quad   \mathrm{on} \quad \gammasN\;,
\end{align}
where, $\nabla_{\bm{x}}$ is the gradient operator with respect to the current configuration, $\rho^s$ is the density of solid, $\bm{d}^s$ is the displacement of the solid, $\bs{\sigma}^s$ is the Cauchy stress tensor, $\bm{f}^s$ is the body force, $\bm{n}^s$ is the unit outward normal on the boundary $\gamma^s$, $\bar{\bm{d}}^s$ is the specified displacement on the boundary $\gammasD$, and $\bar{\bm{t}}^s$ is specified traction on the boundary $\gammasN$.

The Cauchy stress tensor ($\bs{\sigma}^s$) depends upon the type of material model considered for the solid. In this work, a compressible Neo-Hookean model is used. For comprehensive details on hyperelasticity and finite element formulation in the finite strain regime, we refer to \cite{book-fem-BonetWood, book-fem-ZienkiewiczVol2Ed7}.

\subsection{Equations at the fluid-solid interface}
At the interface $\gamma^{f-s}$ between fluid and solid domains, the kinematic velocity constraint and traction equilibrium need to be satisfied. Under the absence of surface tension, these two conditions can be written mathematically as,
\begin{subequations}  \label{interface-eqs}
\begin{align}
\textrm{Kinematic constraint:} & \qquad \hspace{21.5mm} \bm{v}^f = \bm{v}^s, \label{interface-eq1} \\
\textrm{Traction equilibrium:} & \qquad \bs{\sigma}^f \cdot \bm{n}^f + \bs{\sigma}^s \cdot \bm{n}^s = \bs{0}.  \label{interface-eq2}
\end{align}
\end{subequations}

\section{Weak formulation for the FSI problem} \label{sec-formulation}
Only those important details of the finite element formulation that are relevant to the discussion in the context of the present work are presented. For the comprehensive details of the formulations used, the reader is referred to Kadapa et al. \cite{KadapaCMAME2017rigid,KadapaCMAME2018}.

Using the Nitsche's method \cite{NitscheBCpaper} for enforcing the interface constraints at the fluid-solid interface, the weak formulation for the coupled FSI problem is given as: \\
\noindent
find the fluid velocity $\bm{v}^f$, fluid pressure $p^f$, and the structural velocities $\bm{v}^s$ such that for all weighting functions $\bm{w}^f$, $q^f$, and $\bm{w}^s$:

\begin{align} \label{eqn-weakform-coupled}
B^f(\{ \bm{w}^f,q^f \}, \{ \bm{v}^f, p^f \}) - F^f(\{ \bm{w}^f,q^f \}) + B^s(\bm{w}^s, \bm{v}^s) - F^s(\bm{w}^s) \nonumber \\
+ \gamma_{N_1} \, \Igammafs \left[ \bm{w}^f - \bm{w}^s \right] \cdot \left[ \bm{v}^f - \bm{v}^s \right]  \, d \gamma - \Igammafs \left[ \bm{w}^f - \bm{w}^s \right] \cdot  \left[ \bs{\sigma}^f(\{ \bm{v}^f,p^f \}) \cdot \bm{n}^f \right] \, d \gamma \nonumber \\
 - \, \gamma_{N_2} \, \Igammafs  \left[ \bs{\sigma}^f(\{ \bm{w}^f,q^f \}) \cdot \bm{n}^f \right] \cdot \left[ \bm{v}^f - \bm{v}^s \right] \, d \gamma = 0,
\end{align}
where,
$B^f(\{ \bm{w}^f,q^f \}, \{ \bm{v}^f, p^f \})$ and $F^f(\{ \bm{w}^f,q^f \})$ are the bilinear and linear forms for the fluid problem, and $B^s(\bm{w}^s, \bm{v}^s)$ and $F^s(\bm{w}^s)$ are the respective terms for the solid problem. $B^f$ consists of all the terms corresponding to standard mixed Galerkin formulation ($B^f_{\text{Gal}}$) for incompressible Navier-Stokes, SUPG/PSPG stabilisation ($B^f_{\text{Stab}}$) for circumventing the LBB-condition, and ghost-penalty ($B^f_{\text{GP}}$) for overcoming the numerical difficulties arising due to small cut-cells. The expressions for the standard Galerkin terms are provided below, and we refer to \cite{KadapaCMAME2018} for the detailed expressions of additional terms.

\begin{align}
B^f_{\text{Gal}}(\{ \bm{w}^f,q^f \}, \{ \bm{v}^f,p^f \}) &= \int_{\Omega^f} \bm{w}^f \cdot \rho^f \left[\pderiv{\bm{v}^f}{t} + \bm{v}^f \cdot \nabla \bm{v}^f \right] \, d \Omega
+ \int_{\Omega^f} \mu^f \, \nabla \bm{w}^f : \nabla \bm{v}^f \, d \Omega  \nonumber \\
&- \int_{\Omega^f} \left[ \nabla \cdot \bm{w}^f \right] \, p^f \, d \Omega + \int_{\Omega^f} q^f \, \left[ \nabla \cdot \bm{v}^f \right] \, d \Omega \\
F^f(\{ \bm{w}^f,q^f \}) &= \int_{\Omega^f} \bm{w}^f \cdot \bm{f}^f \, d \Omega + \int_{\Gamma_{f_N}} \bm{w}^f \cdot \bm{h}^f \, d \Gamma
\end{align}

\begin{align}
B^s(\bm{w}^s, \bm{v}^s) &= \int_{\omega^s} \bm{w}^s \cdot \rho^s \pderiv{^2 \bm{d}^s}{t^2} \, d \omega
+ \int_{\omega^s} \nabla_{x} \bm{w}^s : \bs{\sigma}^s(\bm{d}^s) \, d \omega \\
F^s(\bm{v}^s) &= \int_{\omega^s} \bm{w}^s \cdot \bm{f}^s \, d \omega + \int_{\gamma^s_N} \bm{w}^s \cdot \bm{h}^s \, d \gamma
\end{align}

The parameter  $\gamma_{\text{N}_2}$ in equation (\ref{eqn-weakform-coupled}) allows to choose between symmetric Nitsche's method ($\gamma_{\text{N}_2} = 1$) and unsymmetric Nitsche's method ($\gamma_{\text{N}_2} = -1$). The parameter $\gamma_{\text{N}_1} (\geq 0)$ is the penalty parameter that controls the accuracy of imposition of the velocity constraint at the interface; the higher the value of $\gamma_{N_1}$, the higher the accuracy. Although the penalty-free version with the unsymmetric version is stable, and therefore, is a viable choice, it is not applicable in the context of Robin BCs.

By choosing the test functions appropriately, the weak form for the coupled FSI problem (\ref{eqn-weakform-coupled}) can be decomposed into the weak forms for the fluid and solid sub-problems. The weak form for the fluid problem reads,
\begin{align}
B^f (\{ \bm{w}^f,q^f \}, \{ \bm{v}^f,p^f \}) - F^f(\{ \bm{w}^f,q^f \}) + \gamma_{N_1} \, \Igammafs \bm{w}^f \cdot \left[ \bm{v}^f - \bm{v}^s \right]  \, d \gamma  \nonumber \\
- \Igammafs \bm{w}^f \cdot \left[ \bs{\sigma}^f \cdot \bm{n}^f \right] \, d \gamma - \, \gamma_{N_2} \, \Igammafs \left[ \bs{\sigma}^f \cdot \bm{n}^f \right] \cdot \left[ \bm{v}^f - \bm{v}^s \right] \, d \gamma = 0,
\end{align}
and the weak formulation for the solid sub-problem becomes,
\begin{align} \label{eqn-weakform-solid-2}
B^s(\bm{w}^s, \bm{v}^s) - F^s(\bm{w}^s) = - \Igammafs \bm{w}^s \cdot \left[ \bs{\sigma}^f \cdot \bm{n}^f + \gamma_{\text{N}_1} [\bm{v}^s - \bm{v}^f] \right] \, d \gamma.
\end{align}

\noindent
The term in the big square brackets on the right-hand side of the equation (\ref{eqn-weakform-solid-2}) is a combination of kinematic and traction terms at the fluid-solid interface; therefore, it leads to Robin boundary condition at the interface.

From (\ref{eqn-weakform-solid-2}), we can write the traction term on the solid problem as
\begin{align}
\bs{\sigma}^s \cdot \bm{n}^s = - \bs{\sigma}^f \cdot \bm{n}^f - \gamma_{\text{N}_1} [\bm{v}^s - \bm{v}^f].
\end{align}

\subsection{Spatial and temporal discretisation}
For the integration in the time domain, the backward-Euler scheme (BDF1), which is implicit and first-order accurate in time, is used for both fluid and solid sub-problems.

The fluid problem is solved on a hierarchical b-spline grid using equal-order linear elements for velocity and pressure. The Dirichlet boundary conditions on the fluid problem are enforced weakly using Nitsche's method. In the context of this work, the details of discretisation of the fluid problem are secondary. We refer the reader to \cite{KadapaCMAME2017rigid,KadapaCMAME2018} for the comprehensive details concerning the discretisation of the fluid problem.

For the solid problem, pure displacement formulation with linear quadrilateral elements are employed. By considering the force contribution only due to terms at the fluid-solid interface, and by taking the finite element discretisations as,
\begin{align}
\bm{d}^s &= \mathbf{N}_{s}  \, \mathbf{d}^s, \quad \bm{v}^s = \mathbf{N}_{s} \, \mathbf{v}^s, \quad \bm{w}^s = \mathbf{N}_{s} \, \mathbf{w}^s
\end{align}
the discrete equations for the solid problem (\ref{eqn-weakform-solid-2}) can be written as,
\begin{align} \label{eqn-solid-discrete1}
  \mathbf{M}_{ss} \, \mathbf{a}^{s} + \mathbf{F}^{\mathrm{int}}(\mathbf{d}^{s}) = \mathbf{F}^{\mathrm{ext,1}} + \mathbf{F}^{\mathrm{ext,2}}
\end{align}
where, $\mathbf{M}_{ss}$ is the mass matrix for the solid sub-problem, $\mathbf{a}^{s}$ is the vector of nodal accelerations, $\mathbf{F}^{\mathrm{int}}(\mathbf{d}^{s})$ is the internal force vector, and the force vector contributions $\mathbf{F}^{\mathrm{ext,1}}$ and $\mathbf{F}^{\mathrm{ext,2}}$ from the interface terms are given by,
\begin{align}
\mathbf{F}^{\mathrm{ext,1}} &= - \, \Igammafs \mathbf{N}_s^{\T} \, \left[ \bs{\sigma}^f \cdot \bm{n}^f \right] \, d \gamma, \\
\mathbf{F}^{\mathrm{ext,2}} &= - \, \gamma_{\text{N}_1} \, \Igammafs \mathbf{N}_s^{\T} \, \left[ \bm{v}^s - \bm{v}^f \right] \, d \gamma.
\end{align}

For the temporal discretisation of (\ref{eqn-solid-discrete1}), we have two options for the treatment of solid velocity term in $\mathbf{F}^{\mathrm{ext,2}}$: (i) explicit treatment and (ii) implicit treatment. In both cases, the current configuration of the fluid-solid interface is treated as the one at the previous time step in the staggered scheme and the previous iteration in the iterative approach.

\subsubsection{Explicit treatment of solid velocity}
In this case, the force term in the solid problem is evaluated from the known field variables at the previous iteration. Accordingly, the discrete equations become
\begin{align} \label{eqn-solid-discrete-expl}
  \mathbf{M}_{ss} \, \mathbf{a}^{s} + \mathbf{F}^{\mathrm{int}}(\mathbf{d}^{s}) = - \, \Igammafs \mathbf{N}_s^{\T} \, \left[ \bs{\sigma}^f_{*} \cdot \bm{n}^f \right] \, d \gamma - \, \gamma_{\text{N}_1} \, \Igammafs \mathbf{N}_s^{\T} \, \left[ \bm{v}^s_{*} - \bm{v}^f_{*} \right] \, d \gamma,
\end{align}
where, $()_{*}$ denotes a currently known field. Thus, in the case of explicit treatment, the solid problem is not modified in terms of its mass, damping and stiffness. Since $\bm{v}^s \approx \bm{v}^f$ due to weak imposition of the velocity constraint, the second term on the right hand side of (\ref{eqn-solid-discrete-expl}) can be safely ignored. This approach has been followed in Kadapa et al. \cite{KadapaCMAME2017rigid,KadapaCMAME2018}.

\subsubsection{Implicit treatment of solid velocity}
Implicit treatment of solid velocity is the commonly employed approach for loosely-coupled FSI schemes based on Robin BCs, and such a treatment has been proven to improve the stability of partitioned FSI schemes with respect to added mass instabilities \cite{BurmanCMAME2009,GiganteCMA2021}. In this case, only fluid velocity and fluid pressure are treated explicitly. Accordingly, the discrete equations can be written as
\begin{align} \label{eqn-solid-discrete-impl2}
  \mathbf{M}_{ss} \, \mathbf{a}^{s} + \mathbf{M}_{fs} \, \mathbf{v}^{s} + \mathbf{F}^{\mathrm{int}}(\mathbf{d}^{s}) = - \, \Igammafs \mathbf{N}_s^{\T} \, \left[ \bs{\sigma}^f_{*} \cdot \bm{n}^f \right] \, d \gamma + \, \gamma_{\text{N}_1} \, \Igammafs \mathbf{N}_s^{\T} \, \bm{v}^f_{*} \, d \gamma
\end{align}
where, $\mathbf{M}_{fs}$ is the mass matrix corresponding to the interface terms, and it is given as
\begin{align}
  \mathbf{M}_{fs} = \gamma_{\text{N}_1} \, \Igammafs \mathbf{N}_{s}^{\T} \, \mathbf{N}_{s} \, d \gamma
\end{align}

Compared to equation (\ref{eqn-solid-discrete-expl}), equation (\ref{eqn-solid-discrete-impl2}) has an additional term, $\mathbf{M}_{fs} \, \mathbf{v}^{s}$, which is a function of the penalty parameter $\gamma_{N_1}$. This additional term is a damping term in the context of structural dynamics, and this added damping term is the reason for improved stability of partitioned FSI schemes based on Robin boundary conditions. However, by introducing an additional artificial damping term, the implicit treatment of solid velocity modifies the governing equations and hence the dynamic characteristics of the solid sub-problem at a fundamental level.

To enforce the interface kinematic constraint on the fluid sub-problem accurately and also to achieve improved stability with respect to added-mass, the value of $\gamma_{N_1}$ should be as high as possible. On the other hand, if the value of $\gamma_{N_1}$ is not sufficiently small enough, the added damping significantly alters the fundamental frequencies of the structural model, thereby affecting the response of the structure. Even the use of the optimal value of $\gamma_{N_1}$ does not help as it is always non-zero.

Moreover, since the velocity constraint at the fluid-solid interface is no longer satisfied in the solid sub-problem even in a weak sense in the case of implicit treatment, it introduces a phase difference in the response of the structure. The higher the value of $\gamma_{N_1}$, the higher the phase difference. Therefore, the improved stability of partitioned FSI schemes based on Robin BCs in which the solid velocity is treated implicitly is at the expense of dynamic characteristics of the structure. Although the value of $\gamma_{N_1}$ might not have a considerable impact on the FSI problem in which the final response a steady state, it severely affects the dynamic response of the structure for problems in which the structure undergoes period response, as will be demonstrated using a numerical example. \\

\noindent
\textbf{Remark 1:} For the penalty-free version of Nitsche's method, the damping term in Eq. (\ref{eqn-solid-discrete-impl2}) vanishes; therefore, there is no difference between the two different treatments of solid velocity. In this case, the improved stability of partitioned FSI schemes solely due to the Robin BC will be lost.

\section{Iterative FSI algorithms} \label{sec-algorithm}
The iterative version of the partitioned FSI scheme based on the force predictor \cite{DettmerIJNME2013,KadapaOE2020} that has been recently presented in Dettmer et al. \cite{DettmerIJNME2020} is employed for resolving the coupling between fluid and solid sub-problems. The pseudocode for iterative partitioned schemes is presented in Algorithm \ref{algo-fsi1} and Algorithm \ref{algo-fsi2}, respectively, for the explicit and implicit treatment of solid velocity. The parameter $\beta \in (0,1)$ is a user-defined \textit{relaxation factor}. FSI problems featuring strong added-mass effects require a small value of $\beta$ for the convergence of iterations. We refer the reader to \cite{DettmerIJNME2020} for the comprehensive details on the spectral analysis of the iterative FSI scheme employed in this work.

\begin{algorithm}[H]
\setstretch{2.0}
\caption{Partitioned scheme with explicit treatment of solid velocity} \label{algo-fsi1}
\begin{algorithmic}[1]
%
\State $\mathbf{F}^{s^{(1)}}_{n+1} = \mathbf{F}^{s}_{n}$
%
%
\For{$k=1$ to $k_{\mathrm{max}}$}
\State Solid problem: Solve $\mathbf{M}_{ss} \, \mathbf{a}^{s^{(k)}}_{n+1} + \mathbf{F}^{\mathrm{int}}(\mathbf{d}^{s^{(k)}}_{n+1}) = \mathbf{F}^{s^{(k)}}_{n+1}$ for $\mathbf{d}^{s^{(k)}}_{n+1}$ and $\mathbf{v}^{s^{(k)}}_{n+1}$
\State Update fluid mesh with $\mathbf{d}^{s^{(k)}}_{n+1}$
\State Fluid problem: Solver for $\mathbf{v}^{f^{(k)}}_{n+1}$, $\mathbf{a}^{f^{(k)}}_{n+1}$, $\mathbf{p}^{f^{(k)}}_{n+1}$ with $\mathbf{v}^{s^{(k)}}_{n+1}$
\State Fluid force: $\mathbf{F}^{*}_{n+1} = \Igammafs \mathbf{N}_s^{\T} \, \left[ \bs{\sigma}^{f^{(k)}}_{n+1} \cdot \bm{n}^f \right] \, d \gamma + \, \gamma_{\text{N}_1} \, \Igammafs \mathbf{N}_s^{\T} \, \left[ \bm{v}^{s^{(k)}}_{n+1} - \bm{v}^{f^{(k)}}_{n+1} \right] \, d \gamma
$
\State Interpolate force: $\mathbf{F}^{s^{(k+1)}}_{n+1} = - \beta \, \mathbf{F}^{*}_{n+1} + (1-\beta) \, \mathbf{F}^{s^{(k)}}_{n+1}$
\EndFor
\end{algorithmic}
\end{algorithm}

\begin{algorithm}[H]
\setstretch{2.0}
\caption{Partitioned scheme with implicit treatment of solid velocity} \label{algo-fsi2}
\begin{algorithmic}[1]
%
\State $\mathbf{F}^{s^{(1)}}_{n+1} = \mathbf{F}^{s}_{n}$
%
%
\For{$k=1$ to $k_{\mathrm{max}}$}
\State Solid problem: Solve $\mathbf{M}_{ss} \, \mathbf{a}^{s^{(k)}}_{n+1} + \mathbf{M}_{fs} \, \mathbf{v}^{s^{(k)}}_{n+1} + \mathbf{F}^{\mathrm{int}}(\mathbf{d}^{s^{(k)}}_{n+1}) = \mathbf{F}^{s^{(k)}}_{n+1}$ for $\mathbf{d}^{s^{(k)}}_{n+1}$ and $\mathbf{v}^{s^{(k)}}_{n+1}$
\State Update fluid mesh with $\mathbf{d}^{s^{(k)}}_{n+1}$
\State Fluid problem: Solve for $\mathbf{v}^{f^{(k)}}_{n+1}$, $\mathbf{a}^{f^{(k)}}_{n+1}$, $\mathbf{p}^{f^{(k)}}_{n+1}$ with $\mathbf{v}^{s^{(k)}}_{n+1}$
\State Fluid force: $\mathbf{F}^{*}_{n+1} = \Igammafs \mathbf{N}_s^{\T} \, \left[ \bs{\sigma}^{f^{(k)}}_{n+1} \cdot \bm{n}^f \right] \, d \gamma - \, \gamma_{\text{N}_1} \, \Igammafs \mathbf{N}_s^{\T} \, \bm{v}^{f^{(k)}}_{n+1} \, d \gamma$
\State Interpolate force: $\mathbf{F}^{s^{(k+1)}}_{n+1} = - \beta \, \mathbf{F}^{*}_{n+1} + (1-\beta) \, \mathbf{F}^{s^{(k)}}_{n+1}$
\EndFor
\end{algorithmic}
\end{algorithm}

\section{Numerical examples}  \label{sec-examples}
The performance of the two algorithms discussed in the previous section is illustrated using two numerical examples: (i) a thick beam undergoing a steady response in cross-flow and (ii) a thin beam undergoing a period response in cross-flow. Since the ratio of densities of solid and fluid domains is one in both the examples, they pose significant added-mass instabilities for partitioned FSI approaches, and therefore, ideally suitable for demonstration in this work. The numerical framework adopted in this work is the cut-cell based finite element method that has been studied extensively in \cite{KadapaCMAME2017rigid,KadapaCMAME2018,KadapaJFS2020,KadapaOE2020}. The reader is referred mainly to Kadapa et al. \cite{KadapaCMAME2018,KadapaJFS2020} for the application of the framework to FSI problem with flexible structures.

All the units are in the CGS system. In all the simulations, the unsymmetric version of Nitsche's method is used. The structural response of the beam is modelled using bi-linear (4-noded) continuum elements. A compressible Neo-Hookean model is considered as the material model for the structure. All the simulations start with zero initial conditions.

\subsection{Thick beam undergoing steady response in cross flow} \label{subsec-example1}
\begin{figure}[H]
\centering
 \includegraphics[trim=0mm 0mm 0mm 0mm, clip, scale=1.0]{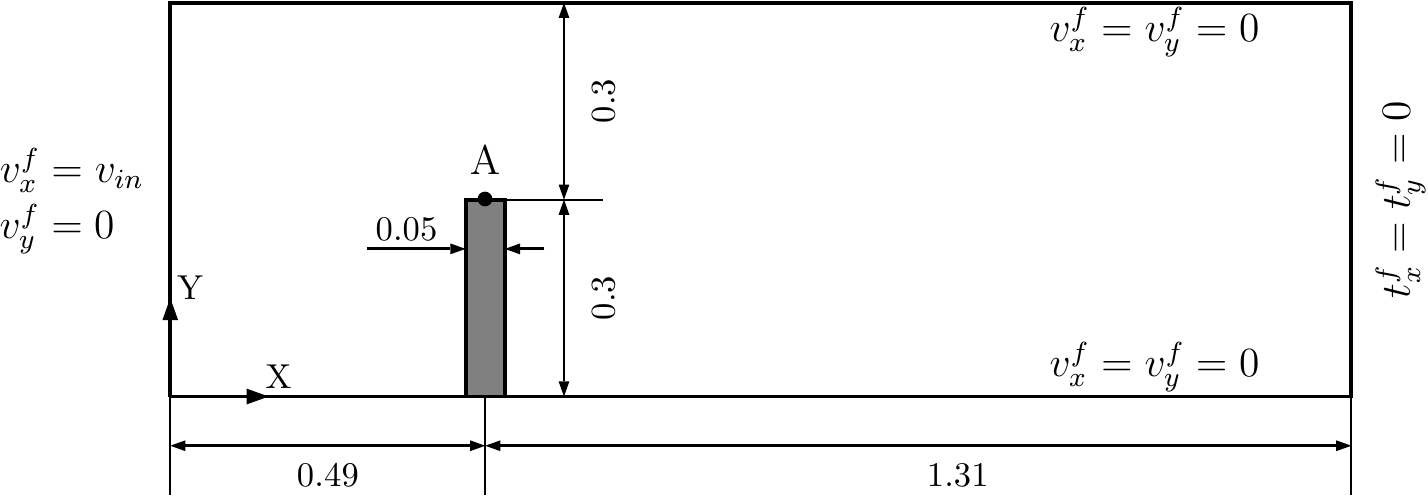}
\caption{Thick beam in cross flow: geometry and boundary conditions. All the dimensions are in cm.}
\label{fig-beam1-geom}
\end{figure}

Previously studied in \cite{KadapaCMAME2018}, this example consists of a thick beam fixed at one end to the bottom wall of a channel. The setup of the problem is depicted in Fig. \ref{fig-beam1-geom}. The parameters are chosen such that a steady-state response ensues. The properties of the fluid are: density, $\rho^f=1.0$, and dynamic viscosity, $\mu^f=0.01$. Material properties of the solid domain are: Young's modulus $E^s=200.0$ and Poisson's ratio $\nu^s=0.3$ is considered. The solid density is taken as $\rho^s=1.0$ so that the density ratio is, $\rho^s/\rho^f=1$. The horizontal velocity profile at the inlet is taken as $v_{in}=\frac{20}{6}y(0.6-y)$. Based on the average velocity at the inlet and length of the beam, the Reynolds number, $Re=6$.

\begin{figure}[H]
\centering
 \includegraphics[trim=0mm 0mm 0mm 0mm, clip, scale=0.38]{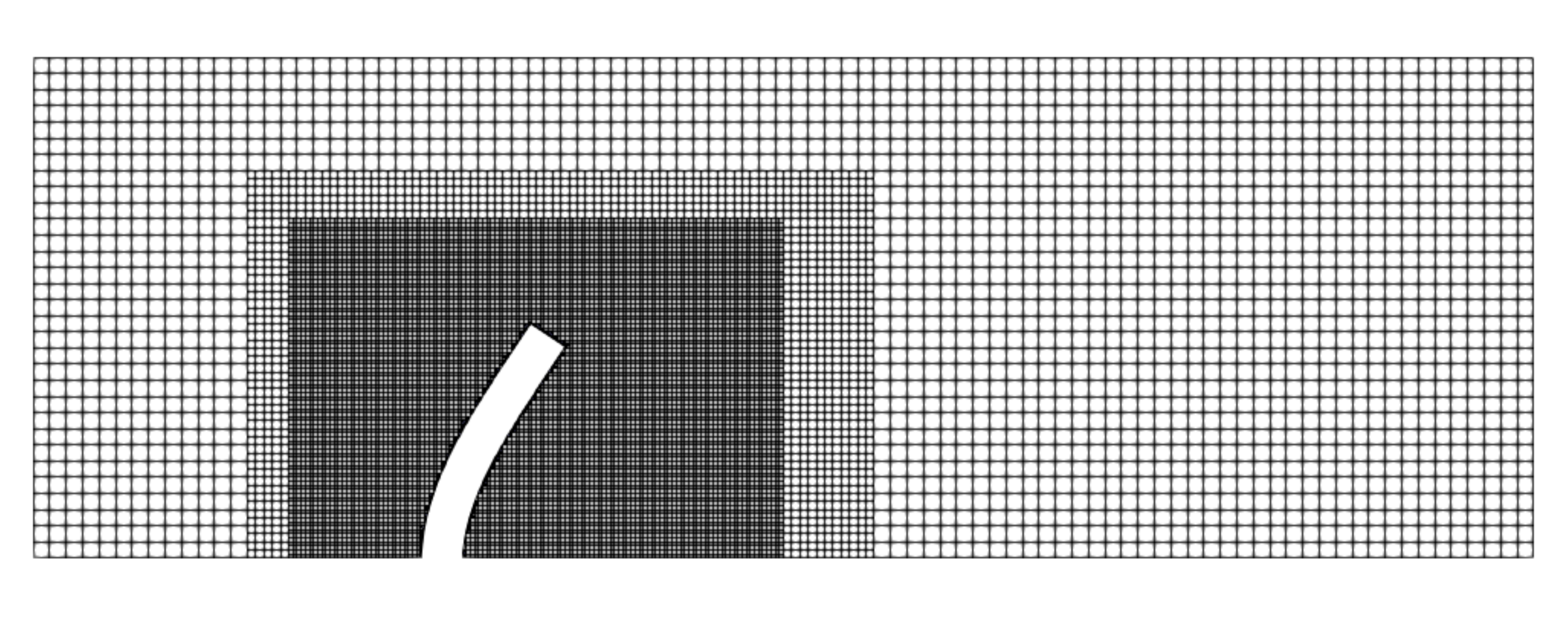}
\caption{Thick beam in cross flow: hierarchical b-spline mesh used for the fluid domain. The configuration corresponds to the solution at $t=10$ s.}
\label{fig-beam1-mesh}
\end{figure}

A two-level hierarchically refined mesh as shown in Fig. \ref{fig-beam1-mesh} is used for the fluid domain, and the beam is discretised with $10 \times 30$ bi-linear elements. The relaxation parameter used in all the simulations is $\beta=0.1$ and the number of iterations ($k_{\mathrm{max}}$) is two. Simulations are carried out using a constant time step size $\Delta t = 0.02$ s until the final time of ten seconds. The results are assessed in terms of the displacement response of point A located at the midpoint of the free end of the beam, see Fig. \ref{fig-beam1-geom}.

\begin{figure}[H]
\centering
\includegraphics[scale=0.7]{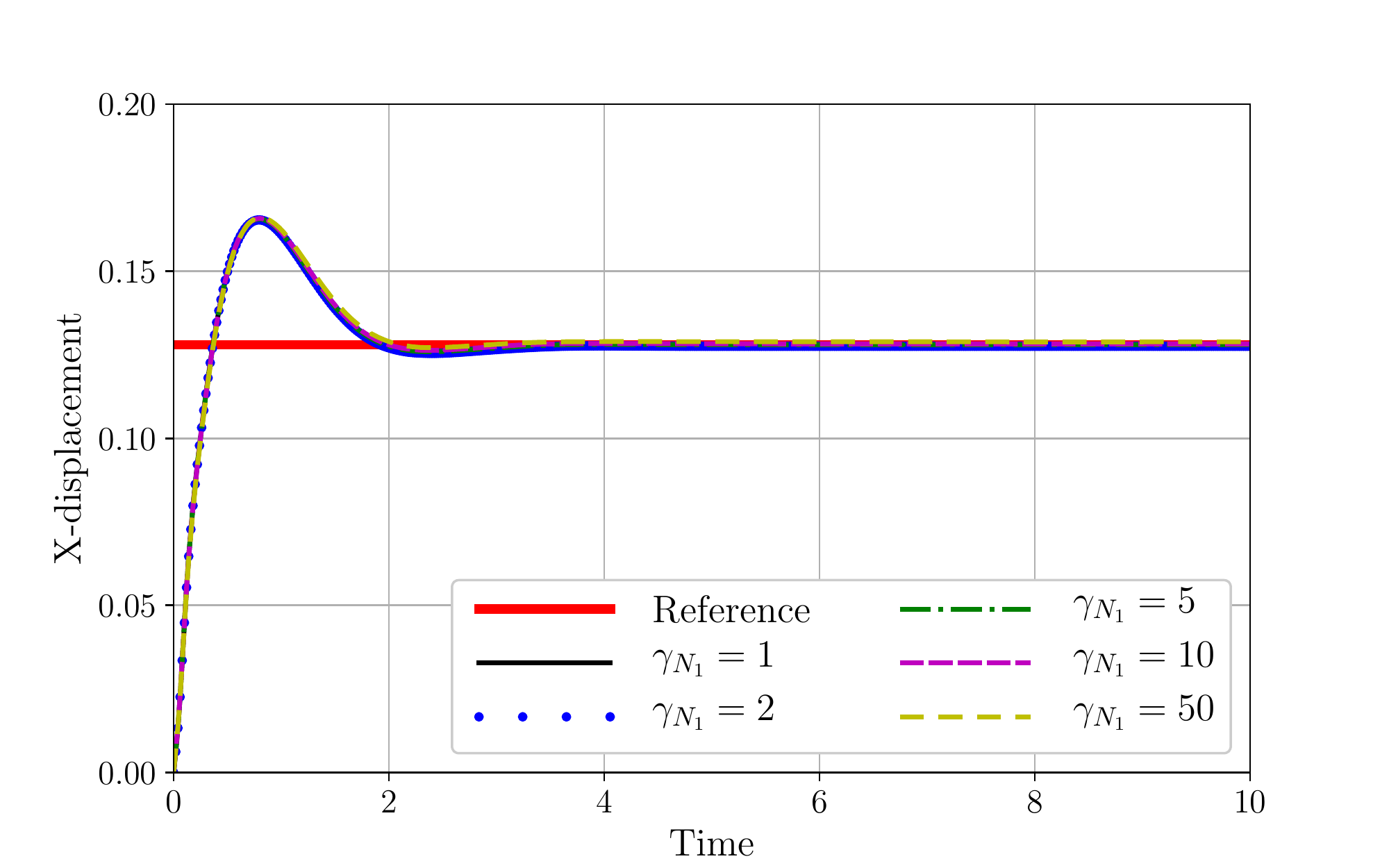}
\caption{Thick beam: displacement response of point A with respect to different values of $\gamma_{N_1}$ using Algorithm 1.}
\label{fig-beam1-graph1}
\end{figure}

\begin{figure}[H]
\centering
\includegraphics[scale=0.7]{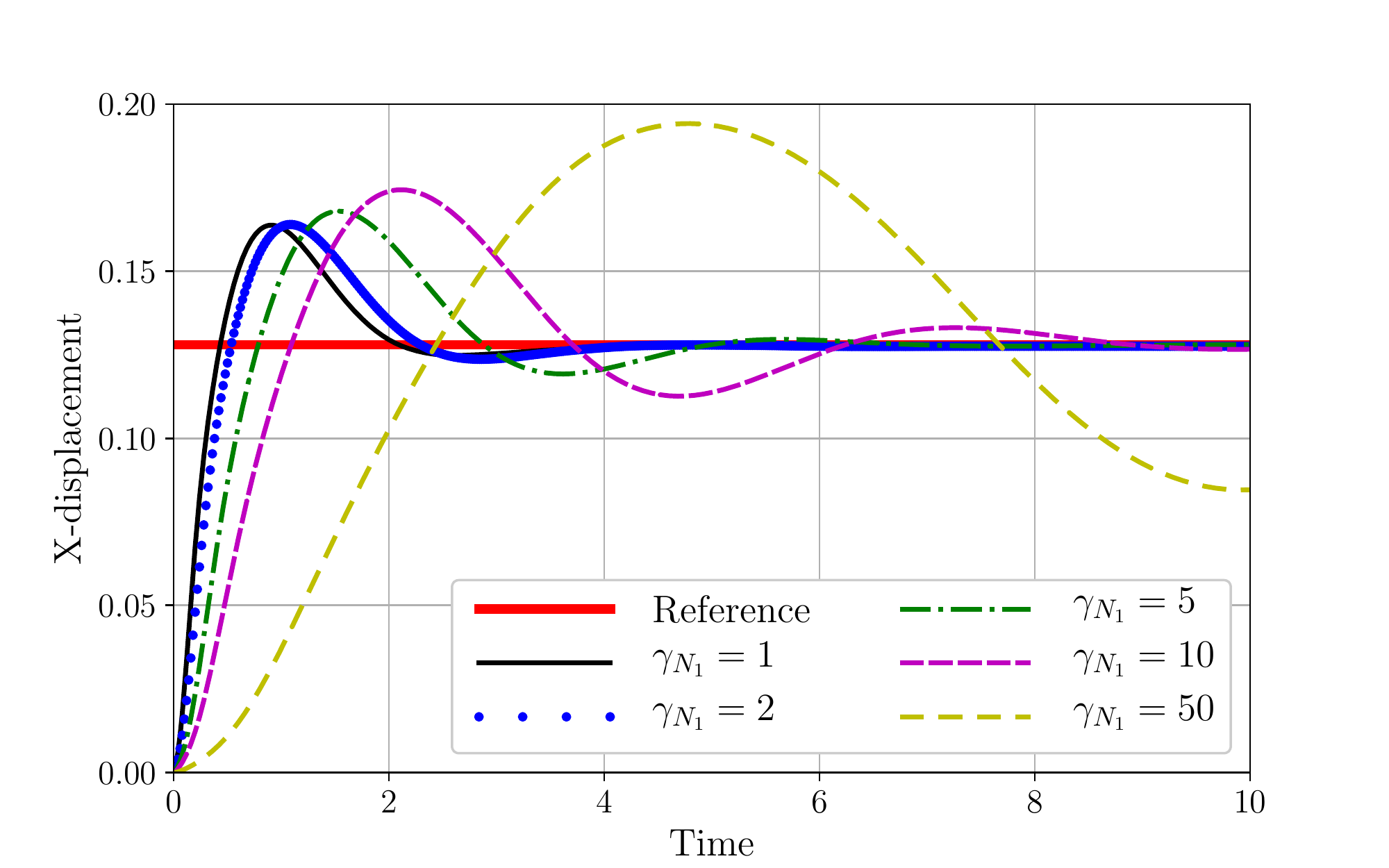}
\caption{Thick beam: displacement response of point A with respect to different values of $\gamma_{N_1}$ using Algorithm 2.}
\label{fig-beam1-graph2}
\end{figure}

The X-displacement of point A obtained for different values of the penalty parameter $\gamma_{N_1}$ using Algorithm 1 is shown in Fig. \ref{fig-beam1-graph1}, and the corresponding graph for Algorithm 2 are presented in Fig. \ref{fig-beam1-graph2}. As shown in Fig. \ref{fig-beam1-graph1}, there is a negligible difference in the response obtained for different values of $\gamma_{N_1}$ using Algorithm 1. However, as shown in Fig. \ref{fig-beam1-graph2}, the response obtained using Algorithm 2 is different for different values of $\gamma_{N_1}$. This response should be expected; as the value of $\gamma_{N_1}$ is increased, the amount of damping increases, which results in reduced natural frequencies of the structure. In other words, increased damping increases the time period of the response of the structure. Additionally, the force contribution due to the fluid velocity term in equation \ref{eqn-solid-discrete-impl2} increases the time lag in response, thus increasing the settling time. The qualitative behaviour of the response shown in Fig. \ref{fig-beam1-graph2} is in good agreement with that observed using a single-degree-of-freedom (SDOF) spring-mass-damper model problem presented in Appendix.

It can also be observed from Fig. \ref{fig-beam1-graph2} that, even though the steady-state can be realised irrespective of the value of $\gamma_{N_1}$ when using Algorithm 2, the computational time required increases with increasing $\gamma_{N_1}$. This is because the dynamic characteristics of the structure are altered significantly when using Algorithm 2. While the altered dynamic characteristics of the structure when using Algorithm 2 might not have a damaging effect for problems with a steady-state solution, the modification due to the added damping term proves to be detrimental for FSI problems involving periodic response as demonstrated in the following example.


\begin{figure}[H]
\centering
\includegraphics[scale=0.5]{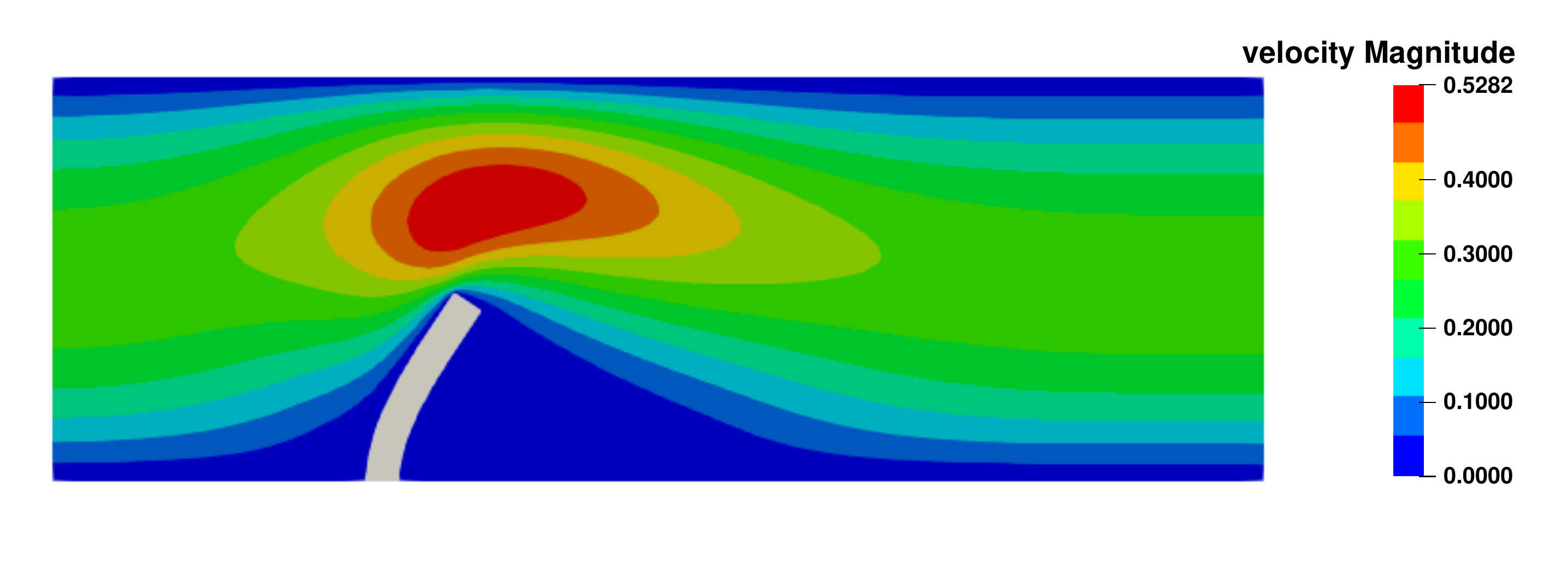}
\caption{Thick beam: contour plot of velocity magnitude at $t=10$ s.}
\label{fig-beam1-contour1}
\end{figure}

\subsection{Thin beam undergoing periodic motion in cross flow}  \label{subsec-example2}
\begin{figure}[H]
\centering
 \includegraphics[clip, scale=1.0]{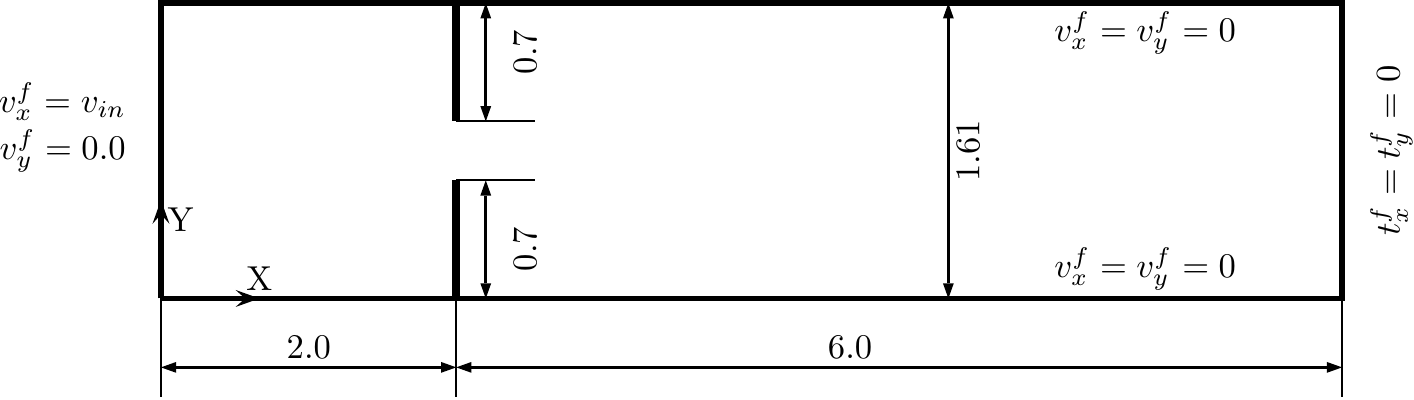}
 \caption{Heart valve: geometry and boundary conditions. All dimensions are in cm.}
\label{fig-heartvalve-geom}
\end{figure}

Introduced in \cite{GilJCP2010} as an idealised two-dimensional model for fluid-structure interaction in the mitral valve, this example is one of the benchmarks used to validate the numerical framework for FSI. The geometry of the problem, as shown in Fig. \ref{fig-heartvalve-geom}, consists of two leaf valves of equal length fixed to the boundaries of a 2D channel. No-slip boundary conditions are applied on the top and bottom sides of the channel, and the outlet is assumed to be traction-free. The density and viscosity of the fluid are $\rho^f=1.0$ and $\mu^f = 0.1$, respectively. Young's modulus and Poisson's ratio of the leaf, respectively, are $E=5 \times 10^5$ and $\nu=0.4$. The thickness of the valve is $h=0.0212$. The horizontal velocity profile at the inlet is, $v_{in} = 5 \, y(1.61 - y) \, (1.1 + \text{sin}(2 \pi t))$. Due to the periodic variation of inlet velocity, the leaves undergo periodic oscillatory motion.

\begin{figure}[H]
\centering
  \includegraphics[trim=0mm 0mm 0mm 0mm, clip, scale=0.35]{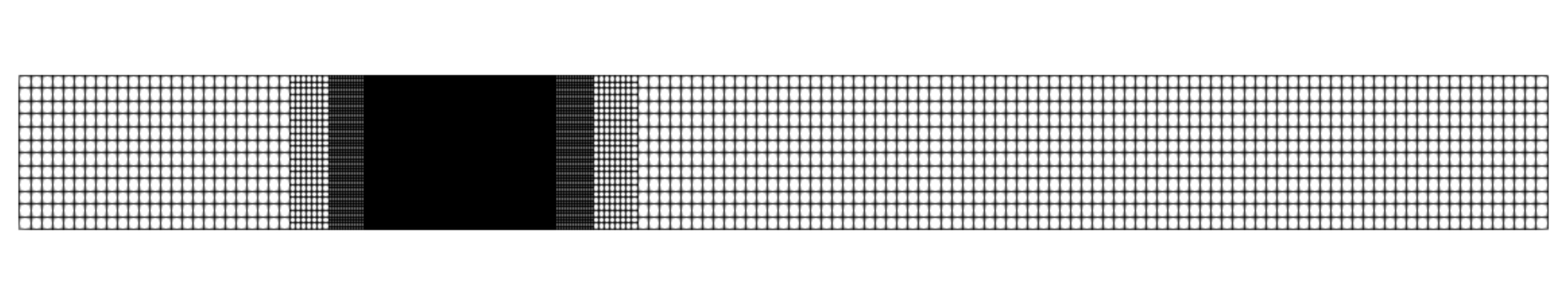}
\caption{Heart valve: hierarchical b-spline mesh for the fluid domain.}
\label{fig-heartvalve-mesh}
\end{figure}

Due to the symmetry of geometry and boundary conditions, the problem is solved over a half portion of the domain. A locally refined hierarchical b-spline mesh as shown in Fig. \ref{fig-heartvalve-mesh} is used for the fluid domain, and it consists of about 50000 DOF. The leaf is discretised using $4 \times 100$ bi-linear quadrilateral elements. The relaxation parameter in all the simulations is $\beta=0.02$ unless stated otherwise.

\begin{figure}[H]
\centering
\includegraphics[scale=0.7]{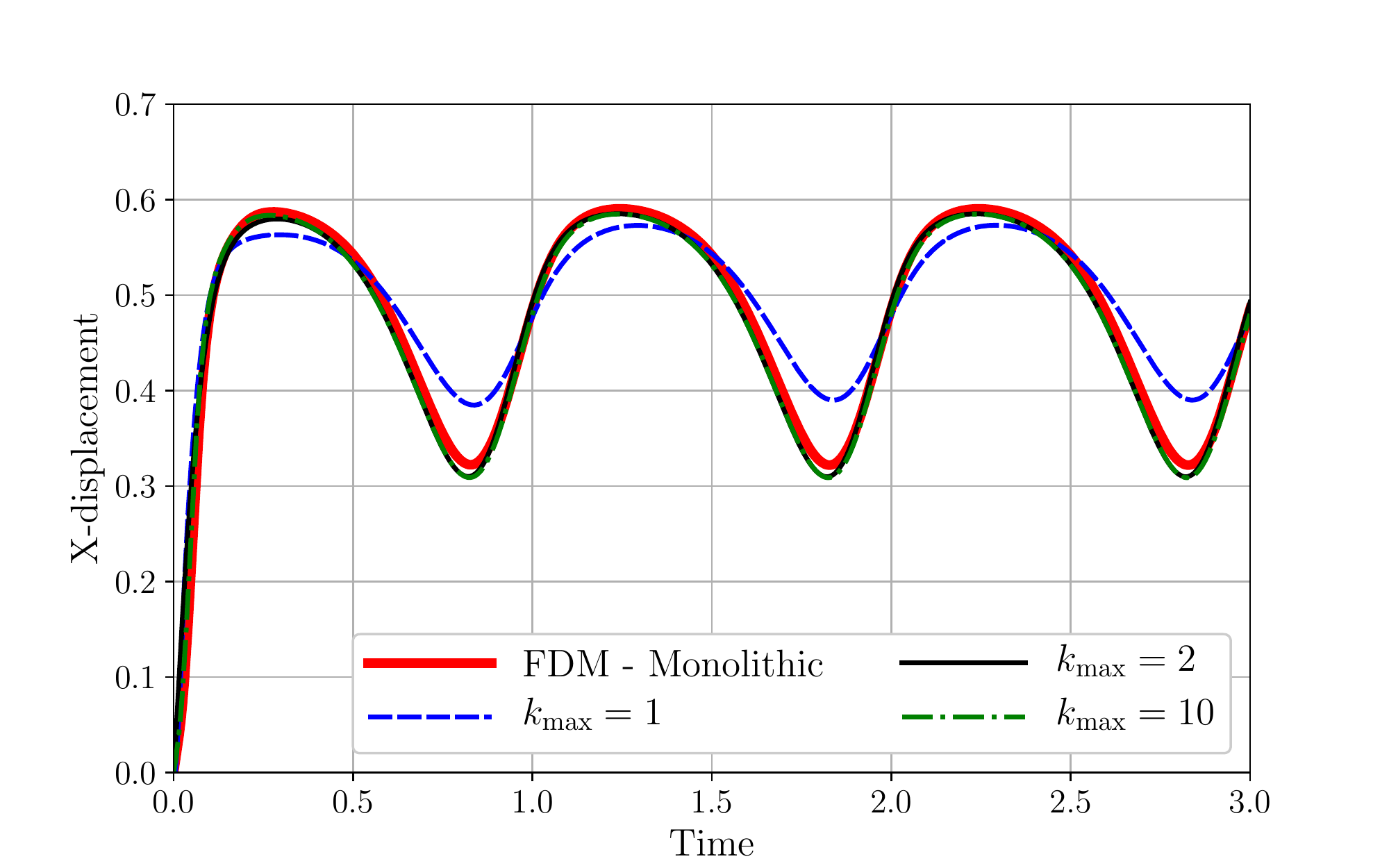}
\caption{Thin beam: X-displacement of point A for different number of maximum iterations ($k_{\mathrm{max}}$) using Algorithm 1 with $\gamma_{N_1}=100$ and $\Delta t=0.005$.}
\label{fig-heartvalve-graph1}
\end{figure}

Based on the numerical experiments presented in Fig. \ref{fig-heartvalve-graph1}, the solution obtained using Algorithm 1 with $k_{\mathrm{max}}=10$, $\gamma_{N_1}=100$ and $\Delta t=0.005$ is treated as the reference solution in subsequent comparisons. The minor difference between the results obtained with the CutFEM approach used in the present work and that obtained with a monolithic scheme based on the fictitious domain method (FDM) \cite{KadapaCMAME2016fsi} are attributed to the fact the solid is modelled as a structural beam element in FDM approach while continuum elements are used in CutFEM approach. Henceforth, $k_{\mathrm{max}}=2$ is used.

\begin{figure}[H]
\centering
\includegraphics[scale=0.7]{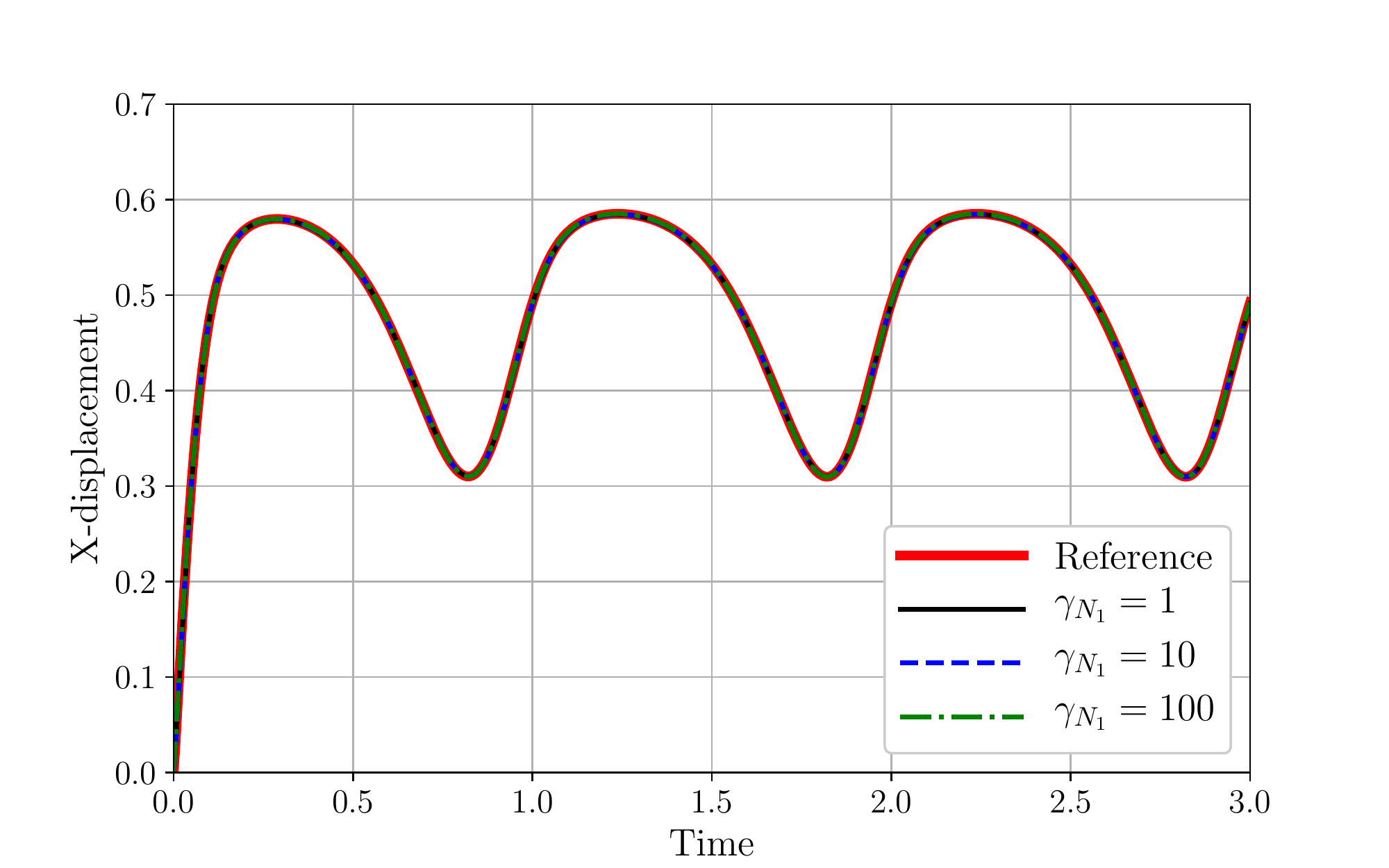}
\caption{Thin beam: X-displacement of point A for different values of $\gamma_{N_1}$ using Algorithm 1 and $\Delta t=0.005$.}
\label{fig-heartvalve-graph-algo1-dt0p005}
\end{figure}

\begin{figure}[H]
\centering
\includegraphics[scale=0.7]{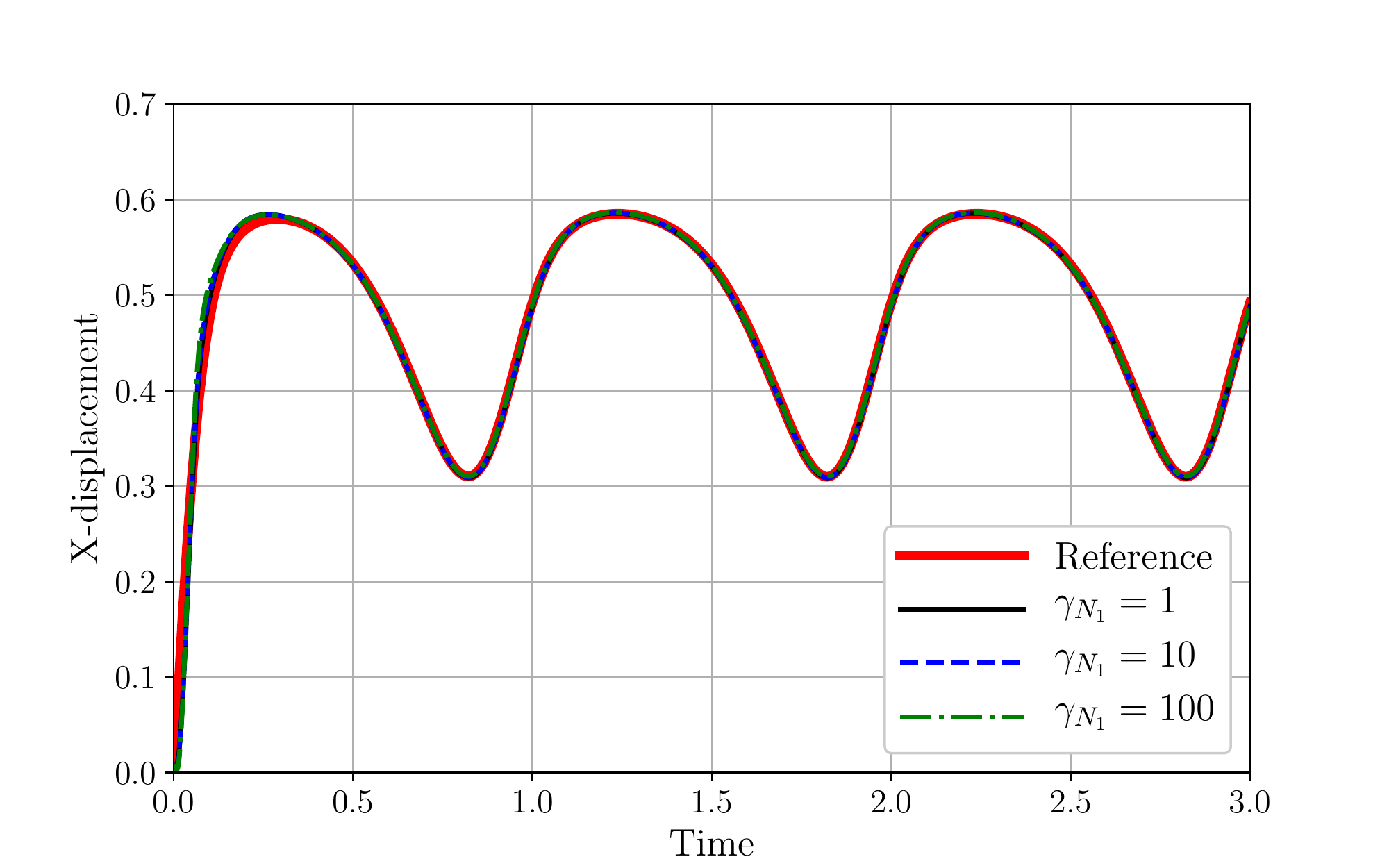}
\caption{Thin beam: X-displacement of point A for different values of $\gamma_{N_1}$ using Algorithm 1 and $\Delta t=0.001$.}
\label{fig-heartvalve-graph-algo1-dt0p001}
\end{figure}

\begin{figure}[H]
\centering
\includegraphics[scale=0.7]{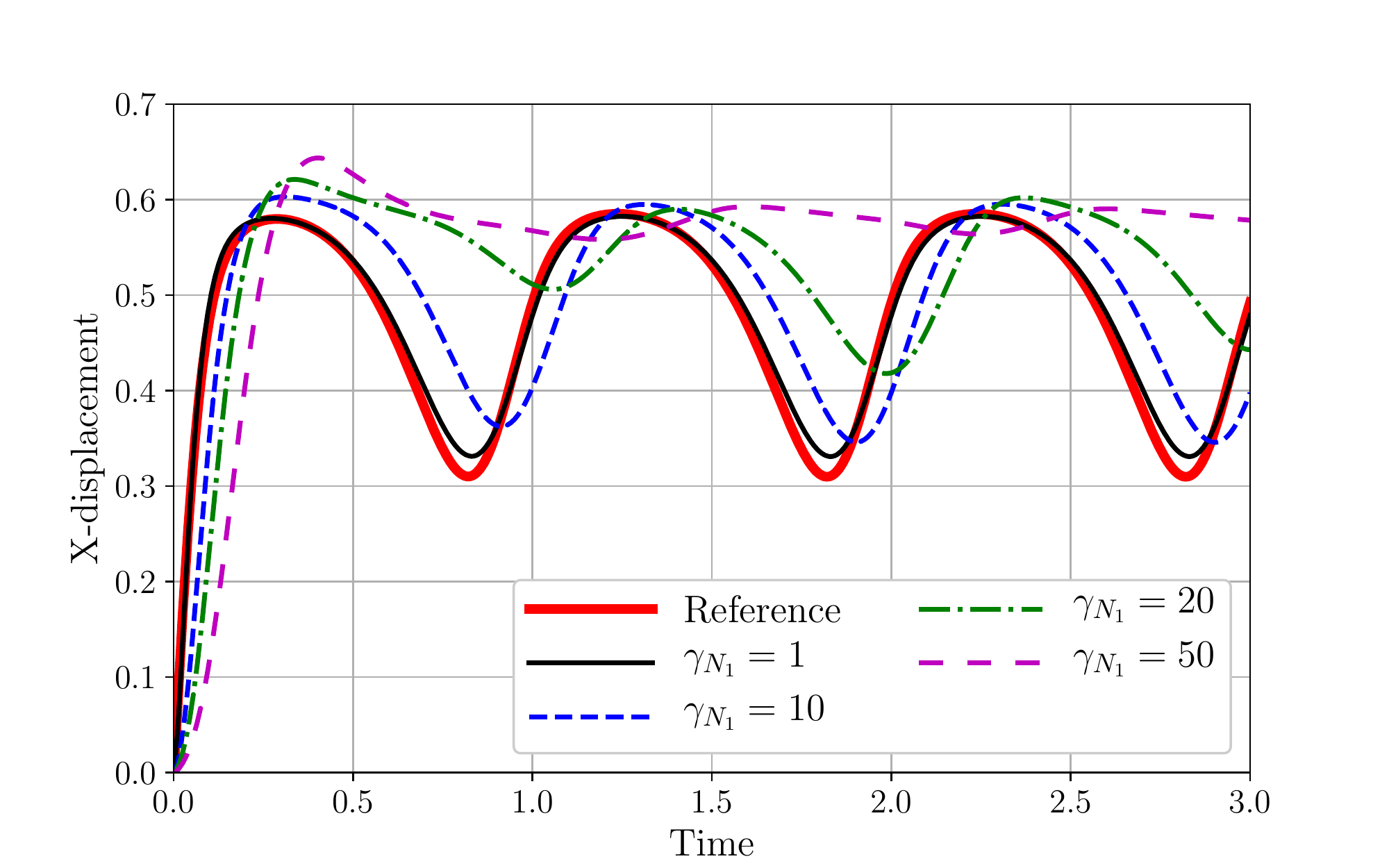}
\caption{Thin beam: X-displacement of point A for different values of $\gamma_{N_1}$ using Algorithm 2 and $\Delta t=0.005$.}
\label{fig-heartvalve-graph-algo2-dt0p005}
\end{figure}

\begin{figure}[H]
\centering
\includegraphics[scale=0.7]{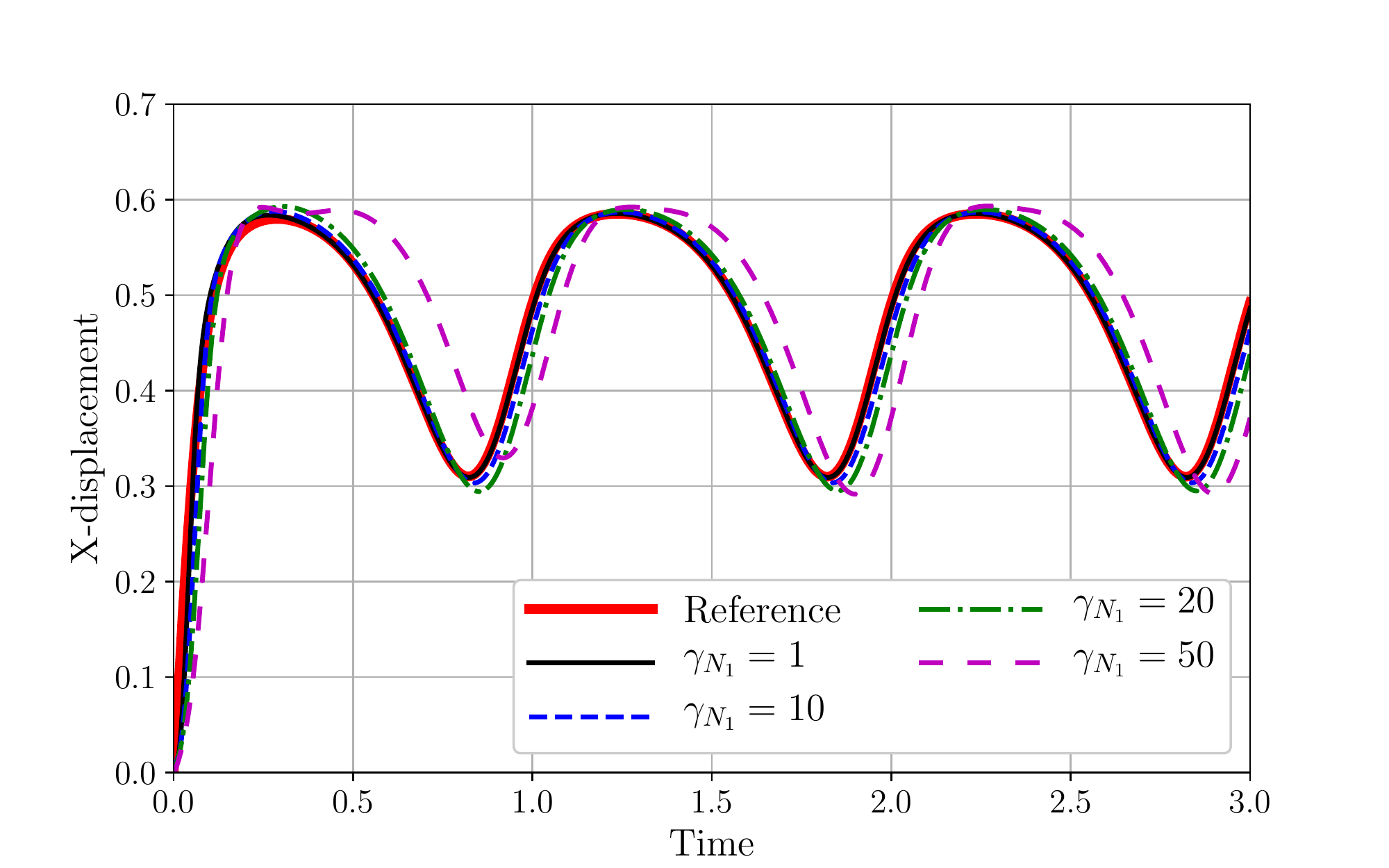}
\caption{Thin beam: X-displacement of point A for different values of $\gamma_{N_1}$ using Algorithm 2 and $\Delta t=0.001$.}
\label{fig-heartvalve-graph-algo2-dt0p001}
\end{figure}

Evolution of X-displacement of midpoint of the free end of a leaf obtained with for different values of $\gamma_{N_1}$ using Algorithm 1 is presented in Fig. \ref{fig-heartvalve-graph-algo1-dt0p005} and Fig. \ref{fig-heartvalve-graph-algo1-dt0p001}, respectively, for $\Delta t=0.005$ and $\Delta t=0.001$. The corresponding graphs for Algorithm 2 are shown in Figs. \ref{fig-heartvalve-graph-algo2-dt0p005} and \ref{fig-heartvalve-graph-algo2-dt0p001}. As shown, Algorithm 1 is robust with respect to both $\gamma_{N_1}$ and $\Delta t$ while Algorithm 2 is not. Results obtained with Algorithm 2 vary significantly depending on the values of $\gamma_{N_1}$ and $\Delta t$. As the value of $\gamma_{N_1}$ is increased, the lag in the response of the structure increases, see Fig. \ref{fig-heartvalve-graph-algo2-dt0p005}; the higher the value of $\gamma_{N_1}$ the higher the lag.  This behaviour is consistent with that observed in the previous example. Although the accuracy of results obtained with Algorithm 2 improves with smaller time step sizes, the error in the response of the structure remains substantial for large values of $\gamma_{N_1}$. Thus, the added artificial damping term in Algorithm 2 significantly influences the response of structures undergoing periodic response. These observations are consistent with the phase mismatch observed in Burman and Fern\'andez \cite{BurmanIJNME2014FSI}, and Gigante and Vergara \cite{Gigante2021,GiganteCMA2021}.

\subsubsection{Study of added-mass}
Further numerical experiments showed that higher values of relaxation parameter ($\beta$) could be used with Algorithm 2, indicating an improved stability limit in Algorithm 2 when compared to Algorithm 1. The maximum permissible value of relaxation parameter ($\beta$) increases with increasing penalty parameter ($\gamma_{N_1}$). This can be attributed to the addition of the artificial damping term in Algorithm 2, which alters the stability of the partitioned FSI algorithm by modifying the dynamic characteristics of the solid problem at a fundamental level. However, the improved stability is at the expense of some essential dynamic characteristics of the structure.


\section{Summary and conclusions} \label{sec-conclusions}
This paper presents some numerical insights into the treatment and performance of partitioned FSI schemes based on Robin boundary conditions. Two algorithms are considered based on the treatment of solid velocity in the Robin boundary condition. The algorithms with the explicit and implicit treatment of solid velocity are denoted, respectively, as Algorithm 1 and Algorithm 2. Two numerical examples are considered to assess the performance of these two algorithms. The observations from numerical results obtained for the two examples can be summarised as follows:
\begin{itemize}
\item Algorithm 1 is robust with respect to the penalty parameter ($\gamma_{N_1}$) and time step size ($\Delta t$), while Algorithm 2 lacks robustness with respect to these two parameters.
\item Provided that the penalty parameter is sufficiently small enough to minimise the effect of added damping term and phase error, Algorithm 2 requires smaller time steps to obtain accurate results compared to Algorithm 1. Therefore, Algorithm 2 is computationally more expensive.
\item This improved stability of Algorithm 2 is due to the added damping term. For large values of $\gamma_{N_1}$, which is essential for the accurate imposition of interface velocity constraint, the artificial damping term and phase error due to violation of interface velocity constraint adversely alters the dynamic characteristics of the structure.
\end{itemize}

The improved stability of partitioned FSI schemes based on Robin BCs is rather due to the implicit treatment of solid velocity term than the boundary condition itself. Present work illustrates that while the implicit treatment of solid velocity in Robin BC certainly improves the stability of the partitioned FSI schemes with respect to added-mass, such treatment proves to be detrimental to the overall dynamic response of the structure. As demonstrated using the numerical example of a thin beam, the improved stability of Algorithm 2 is at the expense of robustness and accuracy, making it less suitable for FSI problems in which the structures undergo periodic response involving significantly large deformations, for example, heart valves or cilia.

We hope that the present work serves as a cautionary tale when using Robin BCs for partitioned FSI schemes. Further research work is necessary towards developing techniques that can improve the stability of partitioned FSI schemes without significantly affecting the dynamic characteristics of individual sub-systems.

\section{References}
\bibliographystyle{unsrt}

\begin{thebibliography}{10}

\bibitem{CausinCMAME2005}
P.~Causin, J.~F. Gerbeau, and F.~Nobile.
\newblock {Added-mass effect in the design of partitioned algorithms for fluid-structure problems}.
\newblock {\em Computer Methods in Applied Mechanics and Engineering},
  194:4506--4527, 2005.

\bibitem{ForsterCMAME2007}
C.~F\"orster, W.~A. Wall, and E.~Ramm.
\newblock {Artificial added mass instabilities in sequential staggered coupling of nonlinear structures and incompressible viscous flows}.
\newblock {\em Computer Methods in Applied Mechanics and Engineering},
  196:1278--1293, 2007.

\bibitem{BrummelenJAM2009}
E.~H. van Brummelen.
\newblock {Added mass effects of compressible and incompressible flows in fluid-structure interaction}.
\newblock {\em Journal of Applied Mechanics}, 76:021206, 2009.

\bibitem{DettmerIJNME2013}
W.~G. Dettmer and D.~Peri\'c.
\newblock {A new staggered scheme for fluid-structure interaction}.
\newblock {\em International Journal for Numerical Methods in Engineering},
  93:1--22, 2013.

\bibitem{BadiaJCP2008}
S.~Badia, F.~Nobile, and C.~Vergara.
\newblock {Fluid-structure partitioned procedures based on Robin transmission conditions}.
\newblock {\em Journal of Computational Physics}, 227:7027--7051, 2008.

\bibitem{NobileSIAMJSC2008}
F.~Nobile and C.~Vergara.
\newblock {An effective fluid-structure interaction formulation for vascular dynamics by generalized Robin conditions}.
\newblock {\em SIAM Journal on Scientific Computing}, 30:731--763, 2008.

\bibitem{BurmanCMAME2009}
E.~Burman and M.~A. Fern\'andez.
\newblock {Stabilization of explicit coupling in fluid-structure interaction involving fluid incompressibility}.
\newblock {\em Computer Methods in Applied Mechanics and Engineering},
  198:766--784, 2009.

\bibitem{GerardoSIAMNA2010}
L.~Gerardo-Giorda, F.~Nobile, and C.~Vergara.
\newblock {Analysis and Optimization of Robin-Robin partitioned procedures in fluid-structure interaction problems}.
\newblock {\em SIAM Journal on Numerical Analysis}, 48(6):2091--2116, 2010.

\bibitem{JaimanCMAME2011}
R.~Jaiman, P.~Geubelle, E.~Loth, and X.~Jiao.
\newblock {Combined interface condition method for unsteady fluid-structure interaction}.
\newblock {\em Computer Methods in Applied Mechanics and Engineering},
  200:27--39, 2011.

\bibitem{FernandezCMAME2013}
M.~A. Fern\'andez, J.~Mullaert, and M.~Vidrascu.
\newblock {Explicit Robin-Neumann schemes for the coupling of incompressible fluids with thin-walled structures}.
\newblock {\em Computer Methods in Applied Mechanics and Engineering},
  267:566--593, 2013.

\bibitem{BurmanIJNME2014FSI}
E.~Burman and M.~Fern\'andez.
\newblock {Explicit strategies for incompressible fluid-structure interaction problems: Nitsche type mortaring versus Robin-Robin coupling}.
\newblock {\em International Journal for Numerical Methods in Engineering},
  97:739--758, 2014.

\bibitem{KadapaOE2020}
C.~Kadapa.
\newblock {A second-order accurate non-intrusive staggered scheme for the interaction of ultra-lightweight rigid bodies with fluid flow}.
\newblock {\em Ocean Engineering}, 217:107940, 2020.

\bibitem{BrummelenIJNMF2011}
E.~H. van Brummelen.
\newblock {Partitioned iterative solution methods for fluid-structure interaction}.
\newblock {\em International Journal for Numerical Methods in Fluids}, 65:3--7,
  2011.

\bibitem{BanksJCP2014part1}
J.~W. Banks, W.~D. Henshaw, and D.~W. Schwendeman.
\newblock {An analysis of a new stable partitioned algorithm for FSI problems. Part I: Incompressible flow and elastic solids}.
\newblock {\em Journal of Computational Physics}, 269:108--137, 2014.

\bibitem{BanksJCP2014part2}
J.~W. Banks, W.~D. Henshaw, and D.~W. Schwendeman.
\newblock {An analysisof a new stable partitioned algorithm for FSI problems. Part II: Incompressible flow and structural shells}.
\newblock {\em Journal of Computational Physics}, 268:399--416, 2014.

\bibitem{SerinoJCP2019}
D.~A. Serino, J.~W. Banks, W.~D. Henshaw, and D.~W. Schwendeman.
\newblock {A stable added-mass partitioned (AMP) algorithm for elastic solids and incompressible flow}.
\newblock {\em Journal of Computational Physics}, 399:108923, 2019.

\bibitem{GiganteCMA2021}
G.~Gigante and C.~Vergara.
\newblock {On the stability of a loosely-coupled scheme based on a Robin interface condition for fluid-structure interaction}.
\newblock {\em Computers and Mathematics with Applications}, 96:109--119, 2021.

\bibitem{RubergCMAME2012}
T.~R\"uberg and F.~Cirak.
\newblock {Subdivision-stabilised immersed b-spline finite elements for moving boundary flows}.
\newblock {\em Computer Methods in Applied Mechanics and Engineering},
  209-212:266--283, 2012.

\bibitem{RubergIJNMF2014}
T.~R\"uberg and F.~Cirak.
\newblock {A fixed-grid b-spline finite element technique for fluid-structure interaction}.
\newblock {\em International Journal for Numerical Methods in Fluids},
  74:623--660, 2014.

\bibitem{DettmerCMAME2016}
W.~G. Dettmer, C.~Kadapa, and D.~Peri\'c.
\newblock {A stabilised immersed boundary method on hierarchical b-spline grids}.
\newblock {\em Computer Methods in Applied Mechanics and Engineering},
  311:415--437, 2016.

\bibitem{KadapaCMAME2017rigid}
C.~Kadapa, W.~G. Dettmer, and D.~Peri\'c.
\newblock {A stabilised immersed boundary method on hierarchical b-spline grids for fluid-rigid body interaction with solid-solid contact}.
\newblock {\em Computer Methods in Applied Mechanics and Engineering},
  318:242--269, 2017.

\bibitem{KadapaCMAME2018}
C.~Kadapa, W.~G. Dettmer, and D.~Peri\'c.
\newblock {A stabilised immersed framework on hierarchical b-spline grids for fluid-flexible structure interaction with solid-solid contact}.
\newblock {\em Computer Methods in Applied Mechanics and Engineering},
  335:472--489, 2018.

\bibitem{book-fem-BonetWood}
J.~Bonet and R.~D. Wood.
\newblock {\em {Nonlinear continuum mechanics for finite element analysis}}.
\newblock Cambridge University Press, Cambridge, 1997.

\bibitem{book-fem-ZienkiewiczVol2Ed7}
O.~C. Zienkiewicz, R.~L. Taylor, and D.~D Fox.
\newblock {\em {The Finite Element Method for Solid and Structural Mechanics}}.
\newblock Butterworth and Heinemann, seventh edition, 2014.

\bibitem{NitscheBCpaper}
J.~A. Nitsche.
\newblock {\"Uber ein Variationsprinzip zur L\"osung von Dirichlet-Problemen bei Verwendung von Teilr\"aumen, die keinen Randbedingungen unterworfen sind}.
\newblock {\em Abhandlungen aus dem Mathematischen Seminar der Universität
  Hamburg}, 36:9--15, 1970-1971.

\bibitem{DettmerIJNME2020}
W.~G. Dettmer, A.~Lovri\'c, C.~Kadapa, and D.~Peri\'c.
\newblock {New iterative and staggered solution schemes for incompressible fluid-structure interaction based on Dirichlet‐Neumann coupling}.
\newblock {\em International Journal for Numerical Methods in Engineering},
  2020.

\bibitem{KadapaJFS2020}
C.~Kadapa, W.~G. Dettmer, and D.~Peri\'c.
\newblock {Accurate iteration-free mixed-stabilised formulation for laminar incompressible Navier-Stokes: Applications to fluid-structure interaction}.
\newblock {\em Journal of Fluids and Structures}, 97:103077, 2020.

\bibitem{GilJCP2010}
A.~J. Gil, A.~A. Carre\~no, J.~Bonet, and O.~Hassan.
\newblock {The immersed structural potential method for haemodynamic applications}.
\newblock {\em Journal of Computational Physics}, 229:8613--8641, 2010.

\bibitem{KadapaCMAME2016fsi}
C.~Kadapa, W.~G. Dettmer, and D.~Peri\'c.
\newblock {A fictitious domain/distributed Lagrange multiplier based fluid-structure interaction scheme with hierarchical B-Spline grids}.
\newblock {\em Computer Methods in Applied Mechanics and Engineering},
  301:1--27, 2016.

\bibitem{Gigante2021}
G.~Gigante and C.~Vergara.
\newblock {On the choice of interface parameters in Robin-Robin loosely coupled schemes for fluid-structure interaction}.
\newblock Technical Report 18/2021, Polytechnic University of Milan, 2021.

\end{thebibliography}

\section*{Appendix: Model problem}
A single degree of freedom (SDOF) linear spring-mass-damper system is considered to gain insights into the performance characteristics. Granted that the SDOF model problem might not capture the exact numerical response of the FSI problem with a flexible structure, but it does offer valuable insights into the qualitative response of the structure when using Algorithm 2.

For the case of an FSI problem in which the solid reaches a steady-state, the fluid force ($\mathbf{F}^{\mathrm{ext,1}}_{n}$) can be assumed to be a constant. Noting that $\bm{v}^{f}_{n}=\bm{v}^{s}_n$, equation (\ref{eqn-solid-discrete-impl2}) for the SDOF model problem can be written as,
\begin{align*}
m_{ss} \, a^s_{n+1} + (c+m_{fs}) \, v^{s}_{n+1} + k \, d^{s}_{n+1} = F^{s}_{n+1} + m_{fs} \, v^{s}_{n}.
\end{align*}

The following values are considered to study the response.
\begin{align*}
m_{ss} = 1.0; \quad c = 2.5; \quad k = 10; \quad F^{s}_{n+1} = 1.282.
\end{align*}

\begin{figure}[H]
\centering
\includegraphics[scale=0.7]{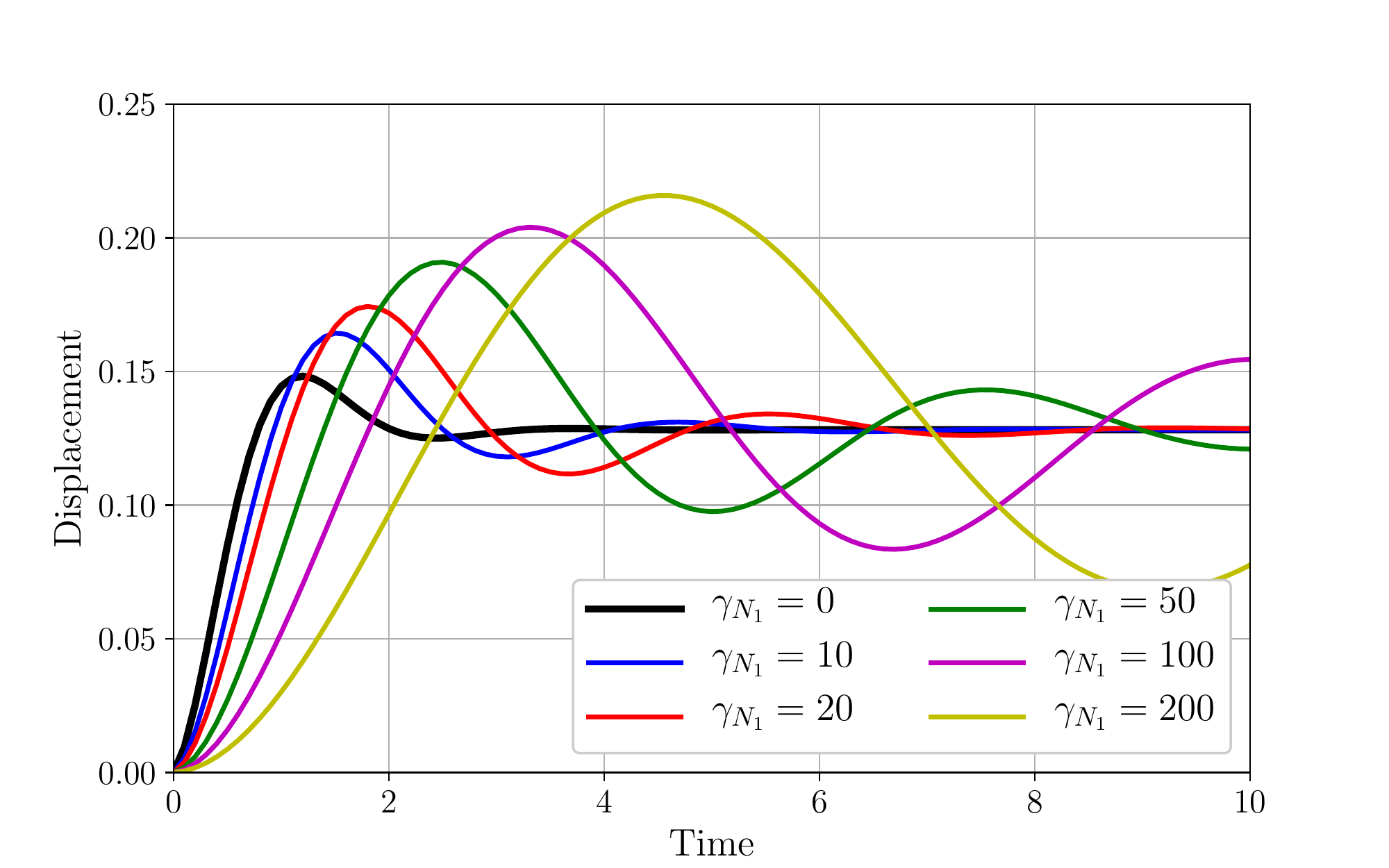}
\caption{Model problem: displacement response with respect to different values of $m_{fs}=\gamma_{N_1}$.}
\label{fig-model-problem}
\end{figure}

The parameters are chosen so that the model problem qualitatively mimics the steady-state response of the structure in the first numerical example. For simplicity, it is assumed that $m_{fs}=\gamma_{N_1}$, and the response of the SDOF model problem is studied for different values of $\gamma_{N_1}$.

The displacement response of the model problem, as shown in Fig. \ref{fig-model-problem}, illustrates that as the value of $\gamma_{N_1}$ is increased, the settling time of the structure increases. The response of the model problem shown in Fig. \ref{fig-model-problem} demonstrates the correctness of the qualitative response of the structure in the example presented in Section \ref{subsec-example1}. Additionally, it proves that if the value of $\gamma_{N_1}$ is not sufficiently small then the added damping term in Algorithm 2 has the potential to affect the response of the structure to a great extent.

\end{document}